\documentclass[9pt,twocolumn,twoside]{opticajnl}
\journal{opticajournal} % use for journal or Optica Open submissions

\setboolean{shortarticle}{false}
\title{Single-shot spectral-encoded waveform reconstruction through probabilistic inversion}

\author[1]{Minsoo Kang}
\author[1,2,*]{Thomas C. Underwood}

\affil[1]{Department of Aerospace Engineering and Engineering Mechanics, University of Texas at Austin, Austin, Texas, 78712, United States}

\affil[2]{Texas Materials Institute, University of Texas at Austin, Austin, Texas, 78712, United States}

\affil[*]{thomas.underwood@utexas.edu}

\begin{abstract}
Spectral encoding enables single-shot measurements of ultrafast transients by mapping temporal information onto the spectrum of a chirped probe. This encoding allows dynamics to be recorded that are beyond the response limits of conventional electronic detectors. However, because the measurements record only spectral intensity, the phase of encoded signals is lost, and dispersion in the detection process introduces waveform distortions that complicate reconstruction and quantitative interpretation of spectra. In single-shot terahertz time-domain spectroscopy (THz-TDS), these distortions manifest as a tradeoff between temporal resolution and the measurement window of signals and can produce spectral null frequencies that limit the recoverable THz bandwidth. To address this challenge, a Bayesian inversion framework is developed to recover the underlying waveform from the squared spectral observable by inferring the THz field, the modulation coefficient, and a low-dimensional empirical parameterization of the probe spectrum jointly, while a Gaussian process (GP) prior regularizes the waveform. The framework is validated using single-shot THz-TDS experiments spanning two probe spectral profiles and three chirp conditions with {\unboldmath $\alpha$} ranging from 14.5 to 40 {\unboldmath $\mathrm{ps}^{-2}$}. Across all cases, the inversion reconstructs both the time-domain waveform and spectral null frequency structure within the credible interval of a delay-line reference measurement. These results establish a pathway to eliminate penalties that are associated with the detection process in spectral encoding methods without adding additional optics or alignment complexity.
\end{abstract}

\setboolean{displaycopyright}{false} % Do not include copyright or licensing information in submission.

\begin{document}

\maketitle

\section{Introduction}
Many transport, reaction, and relaxation phenomena are governed by ultrafast nonequilibrium processes that redistribute energy, charge, and momentum prior to equilibration. Ultrafast laser systems employing femtosecond pulses at high repetition rates have enabled pump-probe diagnostics that resolve these transient phenomena, including carrier transport in semiconductors, chemical reactions, and plasma dynamics~\cite{stanton1994optical,adak2019observation,wang2020single,weakly2023revealing,zewail2000femtochemistry,tancin2021ultrafast,yang2016terahertz,ilyakov2022terahertz,dulat2024single}. However, a fundamental challenge in such measurements is that electronic detectors cannot record sub-picosecond transients (i.e., electric fields) directly. Instead, the temporal information must be encoded into a measurable quantity that can be registered by a slower detector~\cite{zhang2025ultra,mahjoubfar2017time}. Spectral encoding addresses this challenge by mapping temporal information onto each spectral component of a chirped probe pulse. Each temporal instant is associated with a corresponding wavelength, allowing the full transient response to be captured in a single-shot~\cite{sun1998analysis}. This capability is particularly valuable for low-duty-cycle, irreversible, or shot-to-shot varying phenomena and has led to widespread use of chirped-probe spectral encoding in spatiotemporal imaging, frequency-domain transient absorption spectroscopy, electro-optic spectral decoding of electron bunches, photonic time-stretch systems, and chirped-pulse spectral interferometry~\cite{nallapareddy2026probing,shkrob2004frequency,wilke2002single,berden2004high,couture2023single,chien2000single,kim2002single,patel2020simplified,grace2025single}. However, the mapping between time and wavelength is not exact. Each spectral coordinate samples a finite temporal neighborhood around its stationary point, which introduces distortions into the encoded waveform based on the level of dispersion that is added in the detection process. These distortions impose penalties on the temporal resolution and spectral bandwidth of waveforms, complicate quantitative parameter extraction, and ultimately limit the broader applicability of spectral encoding diagnostics.\par
Single-shot THz-TDS is an application of chirped-probe spectral encoding because it enables acquisition of THz waveforms in fast, irreversible, or nonrepetitive processes where conventional delay-line scanned measurements are impractical. This capability has made single-shot THz-TDS a valuable diagnostic for plasma dynamics, irreversible processes, and electron diffraction~\cite{nallapareddy2026probing,roussel2022phase,saha2026traceable,ofori2025unveiling}. However, the spectral encoding process introduces distortions and a time resolution penalty that complicate quantitative waveform reconstruction and limit measurement fidelity~\cite{sun1998analysis,jiang2000measurement,kang2026robust}. To address this challenge, several studies have formulated waveform recovery as an inverse problem and applied deterministic regularization techniques. One approach employed Tikhonov regularization, in which the waveform is reconstructed by regularizing the forward operator with a parameter selected from singular value decomposition (SVD) analysis~\cite{yellampalle2005algorithm}. Another approach used truncated singular value decomposition (TSVD), where weak singular modes of the forward operator are removed to construct a stable pseudo-inverse solution~\cite{wu2018electro}.\par

These methods demonstrate that waveform recovery is possible, while their practical use remains limited to conditions where the undistorted THz waveform is known~\cite{roussel2022phase}. Since single-shot measurements are limited by noise~\cite{nallapareddy2026probing}, modes with small singular values become difficult to distinguish from noise. These modes correspond to spectral nulls of the measurement and make the reconstruction sensitive to the chosen regularization parameter or truncation rank~\cite{hansen1990truncated,hansen1998rank}. Furthermore, since these methods act on an abstract singular mode space, the regularization cannot be guided by the physical features of the measurement. When the true waveform is unavailable as a reference, this absence of a physical criterion prevents the interpretation of the reconstruction from the measurement alone and requires extra calibration settings.\par

Therefore, most spectral encoding inversion research has focused on hardware based methods~\cite{roussel2022phase}. Supercontinuum-enhanced spectral encoding (SETS) decreased the distortion by adjusting the chirp characteristic of the probe, broadening the probe bandwidth, and increasing the chirp parameter ($\alpha$)~\cite{Nallapareddy2025a}. Phase-diversity electro-optic sampling (DEOS) modified the electro-optic detection geometry to generate two complementary spectral-encoded observables whose spectral nulls occur at different frequencies. These complementary observables provide additional information that mitigates the loss of spectral information near the null frequencies~\cite{roussel2022phase}. Dual-echelon single-shot detection geometrically encoded the THz waveform by mapping the THz time delay axis onto a spatial coordinate of the probe, allowing multiple time points of the waveform to be sampled in a single laser shot~\cite{teo2015invited}. The tomographic method has also been applied to extract probe phase modulation information directly from the probe spectral interferogram measured by a spectrometer~\cite{matlis2010single}. However, these approaches require additional optical hardware beyond the standard spectral encoding setup, such as a nonlinear crystal stage, extra laser lines, and an optical delay setup. These components increase the complexity of the optical system adding additional uncertainties. Also, spatial information is lost in the case using echelon mirrors.\par

These challenges can be addressed without additional hardware by formulating an inversion problem numerically within a Bayesian framework. Within this work, a posterior distribution over the unknown waveform can be constructed from the measurement likelihood and an explicit prior~\cite{kaipio2005statistical,stuart2010inverse}. Unlike regularization methods, the prior can be assigned directly to physical quantities (e.g., probe electric field, THz waveform) so that constraints from the characterized measurement process, such as spectral nulls, can guide the inversion when the target waveform is not known a priori. The resulting posterior can provide intrinsic uncertainty estimates for input parameters, including the recovered waveform. The Bayesian formulation is also compatible with data-driven extensions, including parameter estimation or machine-learning-based priors that are increasingly used in ultrafast diagnostics~\cite{brown2021bayesian,genty2021machine}.\par

In this work, we demonstrate the first algorithmic inversion of single-shot spectral-encoded waveforms that restores temporal resolution and recovers spectral content which is lost during the measurement process. A Bayesian framework is introduced to account for systematic uncertainties, reduce noise susceptibility, and decrease the need for a reference measurement by using physically informed priors across a wide range of probe conditions. The THz waveform $E_{\mathrm{THz}}$ is treated as the target parameter, together with nuisance parameters including the modulation factor $k$ and low-dimensional chirped-probe parameters. A Gaussian process (GP) prior is assigned to $E_{\mathrm{THz}}$, with its variance set directly by the spectral nulls of the single-shot spectral measurement. Therefore, the prior relies only on the knowledge of the probe beam and does not require a separate reference THz waveform measurement. The posterior is explored by maximum a posteriori (MAP) estimation and the Metropolis-adjusted Langevin algorithm (MALA), and the framework is validated against delay-line reference measurements across six configurations spanning two probe spectral profiles and three chirp conditions, with $\alpha$ from 14.5 to 40~$\mathrm{ps}^{-2}$.\par

\section{Source of Distortion}
Waveform distortion in spectral encoding occurs because the encoded signal is not a one-to-one mapping of the target waveform signal. Spectral encoding is based on the dispersive readout process of a probe pulse, followed by the measurement of its intensity~\cite{nallapareddy2025characterization}. In a pump-probe measurement, the pump-induced target waveform is first imprinted onto a probe as a time-dependent modulation, and the modulated probe $I_m$ is then mapped to a detector coordinate by the readout optics. Several spectral encoding processes, including frequency-domain single-shot transient absorption spectroscopy~\cite{shkrob2004frequency}, single-shot supercontinuum spectral interferometry~\cite{kim2002single}, and chirped-probe coherent anti-Stokes Raman scattering (CARS)~\cite{knutsen2006chirped} can be described as,
\begin{equation}
    I_{\mathrm{m}}(x)
    =
    \left|
        \int_{-\infty}^{\infty}
        K(x,t)E_c(t)S_{\mathrm{m}}(t)\,dt
    \right|^2,
    \label{eq:general_spectral_encoding_forward}
\end{equation}
where $E_c(t)=A_c(t)\exp[-i\alpha t^2]$ is the chirped probe field, $x$ is the readout coordinate, and $K(x,t)=A_K(x,t)\exp[i\psi_K(x,t)]$ is the readout kernel that maps the modulated probe field from time to the measurement coordinate. The sample-dependent modulation $S_m(t)$ can be decomposed into $S_{\mathrm{m}}(t)=S_0(t)+\Delta S(t)$, where $S_0(t)$ is the reference or background response, and $\Delta S(t)$ is the target-induced modulation to be recovered. Thus, the measured spectrum is not a direct measurement of $\Delta S(t)$, but the square magnitude of the
chirped-probe field after propagation through the effective encoding kernel
$K(x,t)E_c(t)$. The integral in \eqref{eq:general_spectral_encoding_forward} becomes nonzero only near $t_{sp}$, where the total phase is stationary,
\begin{equation}
    \Phi(x,t)=\psi_K(x,t)-\alpha t^2,
    \qquad
    \left.
    \frac{\partial \Phi(x,t)}{\partial t}
    \right|_{t=t_{sp}}
    =
    0,
    \label{eq:mapped_time_stationary_phase}
\end{equation}
generating the mapping relation $\tau\equiv x(t_{sp})$. The measurement is then defined from the normalized subtraction between modulated $I_m(\tau)$ and reference $I_c(\tau)$ intensities, $[I_{\mathrm{m}}(\tau)-I_c(\tau)]/I_c(\tau)$. Expanding the integral in \eqref{eq:general_spectral_encoding_forward} near $t_{sp}$ and retaining the leading target-induced modulation $\Delta S(t)$,
\begin{equation}
\begin{split}
    \frac{I_m-I_0}{I_0}(\tau)
    \approx
    &
    2\,\mathrm{Re}\!\left[
        H(\tau)\Delta S(\tau)
    \right] \\
    &+
    \left|
        H(\tau)
        \left(
            \Delta S(\tau)
            +
            \frac{i}{2\kappa}\mathcal{D}_{\Delta S}(\tau)
        \right)
    \right|^2 ,
\end{split}
    \label{eq:generic_distortion}
\end{equation}
where $H(\tau)=C(\tau)A_K(\tau,t_{sp})A_c(\tau)/\tilde{E}_0(\tau)$, $\kappa=\partial_t^2\Phi(\omega,t)|_{t=t_{sp}}$, and $\mathcal{D}_{\Delta S}(\tau)$ collects derivative terms of $A_K(\tau,t)$, $A_c(t)$, and $\Delta S(t)$~(SI Sec. 1.A). The distortion arises from the second term on the right-hand side of \eqref{eq:generic_distortion}, which adds a $\kappa$-dependent error to the linear response of the target $\Delta S(\tau)$. This contaminates $\Delta S(\tau)$ and biases both time domain transient analysis and parameter estimations. Furthermore, because of the squared magnitude of the complex term, the phase of this additional quadrature contribution is lost, making the recovery of $\Delta S(\tau)$ in \eqref{eq:generic_distortion} ill posed. Therefore, the distortion cannot be removed by a simple algebraic correction or recovery, and it produces resolution limits in the extracted signal~\cite{shkrob2004frequency,berden2004high,polli2010effective,couture2023single,nallapareddy2025characterization}.\par
\begin{figure}[!t]
\centering
\includegraphics[width=0.95\linewidth]{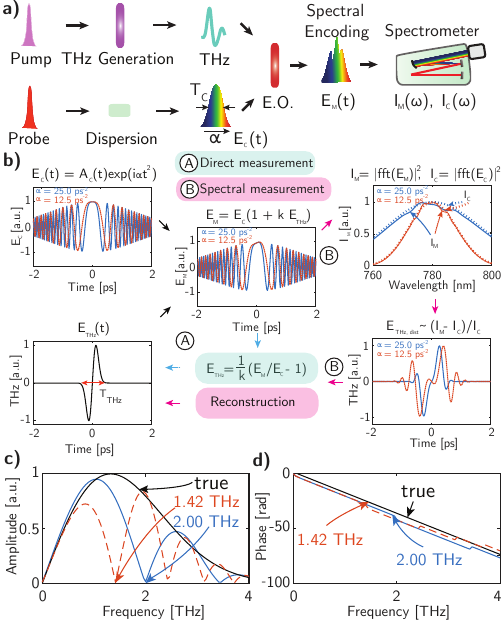}
\caption{The distortion problem in spectral encoding. (a) Physical layout of single-shot THz spectroscopy. (b) Theoretical process: path~A is the direct inversion $E_{\mathrm{THz}} = (E_m/E_c - 1)/k$, and path~B is the spectral measurement mapping $E_{THz}(t)$ on $I_m(\omega)$, obtaining $E_{\mathrm{THz},\mathrm{dist}} \sim (I_m - I_c)/I_c$. Two chirp values ($\alpha = 12.5$ and $25~\mathrm{ps}^{-2}$) are compared on ideal bipolar signal with $T_{THz}$. Distortion and spectral nulls in spectral (c) amplitude and (d) phase.}
\label{fig:Fig1}
\end{figure}
As illustrated in Fig.~\ref{fig:Fig1}(a), the spectral encoding implementation of single-shot THz-TDS proceeds by spatially and temporally overlapping a chirped probe pulse with the generated THz pulse in an electro-optic crystal. Through the Pockels effect, the THz field induces a transient electro-optic modulation of the probe, encoding the THz waveform onto different spectral components of the probe~\cite{sun1998analysis}. The modulated probe is then read out as a single-shot intensity observable, either directly in the frequency domain with a spectrometer or after dispersive mapping into the time domain. In the weak modulation limit, the modulated probe field after the electro-optic crystal can be written as,
\begin{equation}
    E_m(t)
    =
    E_c(t)\left[1+kE_{\mathrm{THz}}(t)\right],
    \label{eq:THz_EO_forward}
\end{equation}
where $E_{\mathrm{THz}}(t)$ is the THz electric field and modulation factor $k\sim\mathcal{O}(10^{-1})$~\cite{sun1998analysis,jiang2000measurement,Nallapareddy2025a,nallapareddy2026probing,saha2026traceable}. If $E_m(t)$ were measured directly, the THz waveform could be recovered algebraically by $E_{\mathrm{THz}} = (E_m/E_c - 1)/k$ (A in Fig~\ref{fig:Fig1}(b)). However, since THz transients typically evolve on picosecond timescales, single-shot measurements require the transient field to be encoded onto a measurable intensity observable such as an intensity spectrum. For THz-TDS, each term in \eqref{eq:generic_distortion} corresponds to $S_0(t)=1,~\Delta S(t) = kE_{\mathrm{THz}}(t),~x = \omega,~K(\omega,t)=e^{i\omega t},~\kappa=2\alpha$. Therefore, the generic distortion turns into (see SI Sec. 1.B),
\begin{equation}
\begin{split}
        &\frac{I_m-I_c}{I_c}(\tau) \approx 2 k E_{THz}(\tau)\\ &+ \left( \frac{k}{4\alpha}\right)^2 \left( E_{\mathrm{THz}}''(\tau) + \frac{2 A_c'(\tau)\, E_{\mathrm{THz}}'(\tau)}{A_c(\tau)} \right)^2.
\end{split}
    \label{eq:THz-TDS_distortion}
\end{equation}
When the chirp rate $\alpha$ becomes small ($\alpha\sim10~\mathrm{ps}^{-2}$), the second term of \eqref{eq:THz-TDS_distortion} rises, introducing distortion to $E_{THz}(\tau)$ that scales with $\alpha$ and its derivatives. This is shown in Fig.~\ref{fig:Fig1}(b), the two chirp values $\alpha = 25~\mathrm{ps}^{-2}$ and $\alpha = 12.5~\mathrm{ps}^{-2}$ illustrate how the measured waveform $E_{\mathrm{THz,dist}}$ deviates with a ringing feature from the true THz $E_{\mathrm{THz}}$ as $\alpha$ varies. The effect of waveform distortion in the spectral domain is shown in Fig.~\ref{fig:Fig1}(c) and (d). Assuming that $A_c(t)\approx1$ and expanding higher order terms in \eqref{eq:THz-TDS_distortion}, the transfer function can be written as $H(\omega) = 2k\mathrm{cos}(\frac{\omega^2}{4 \alpha})$. The zeros of this transfer function introduce null frequencies at $f_{\mathrm{null}} = \sqrt{(2n{-}1)\alpha/2\pi},~n = 1, 2,\ldots$. These null frequencies give rise to the distortion by suppressing the corresponding spectral components and introducing phase jumps through sign changes across the nulls. As a result, quantitative spectroscopy and broadband THz imaging are limited to the usable
bandwidth below the first null $f_{null} = \sqrt{\frac{\alpha}{2\pi}}$, which is 2.00~THz at $\alpha = 25~\mathrm{ps}^{-2}$ and 1.41~THz at $\alpha = 12.5~\mathrm{ps}^{-2}$.~\cite{Nallapareddy2025a, roussel2022phase, nallapareddy2025characterization, kang2026robust}.\par

\section{Bayesian Inversion Framework}
\begin{figure}[!t]
\centering
\includegraphics[width=0.95\linewidth]{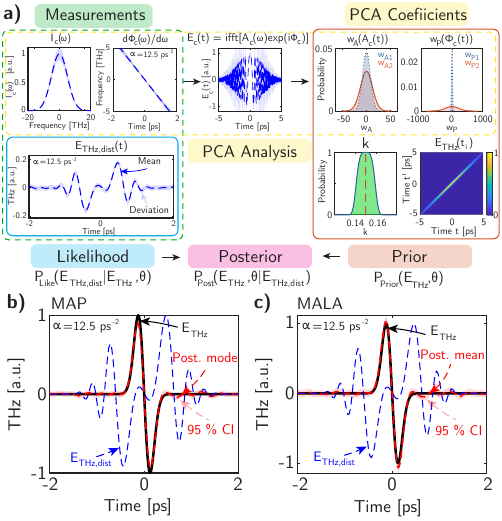}
\caption{(a) Schematic of the Bayesian inversion algorithm using a Gaussian process (GP) prior. The workflow spans from measured data to principal component analysis (PCA), definitions of the parameters ($E_{\mathrm{THz}}(t_i)$ and $\theta=\{k, w_A, w_P\}$), likelihood $P_{\mathrm{Like}}(E_{\mathrm{THz,dist}}|E_{\mathrm{THz}}, \theta)$, prior $P_{\mathrm{Prior}}(E_{\mathrm{THz}}, \theta)$, and target posterior $P_{\mathrm{Post}}(E_{\mathrm{THz}}, \theta|E_{\mathrm{THz,dist}})$. Inversion results obtained by (b) maximum a posteriori (MAP) and (c) Metropolis-adjusted Langevin algorithm (MALA), applied to an ideal bipolar signal ($T_{\mathrm{THz}} = 0.2~\mathrm{ps}$) overlaid with $E_{\mathrm{THz,dist}}$ for $\alpha = 12.5~\mathrm{ps}^{-2}$.}
\label{fig:Fig2}
\end{figure}

The Bayesian inversion framework developed in this work infers the THz waveform $E_{\mathrm{THz}}(t_i)$ together with nuisance parameters $\theta=\{k,\mathbf{w}_A,\mathbf{w}_P\}$ from the measured single-shot distorted field $E_{\mathrm{THz,dist}}$ (Fig.~\ref{fig:Fig2}(a)). The workflow begins by characterizing the chirped probe field $E_c(t)=A_c(t)\exp[i\phi_c(t)]$ using experimental shot-to-shot measurements of the unmodulated probe spectrum $I_c(\omega)$ and its corresponding group-delay profile $\frac{d\phi_c(\omega)}{d\omega}$. The group delay is calibrated by scanning the probe with THz, adjusting the arrival time at the electro-optic crystal, and tracking the THz peak shift. The square root of $I_c(\omega)$ gives the amplitude $A_c(\omega)$, and the calibration curve gives the spectral phase $\phi_c(\omega)$. Combining these two components yields the complex field in the frequency domain $\tilde{E}_c(\omega)$. Then, the inverse Fourier transformation on each $\tilde{E}_c(\omega)$ yields an empirical ensemble of complex temporal probe fields, which anchors the prior on $E_c(t)$ (SI Sec. 2.A--E).\par
The probe ensemble is then compressed through principal component analysis (PCA). Shot-to-shot deviations of $A_c(t)$ and $\phi_c(t)$ from their ensemble means are projected onto orthonormal bases, with latent weights $\mathbf{w}_A$ and $\mathbf{w}_P$ modeled by empirical zero-mean Gaussian priors (Fig.~\ref{fig:Fig2}(a)). Only the leading components are retained, up to a cumulative explained variance of 90 \%, keeping two or three weights per variable. The retained components capture the dominant variations, while the discarded components make only minor contributions to the probe field. This truncation reduces the dimension of the variables while preserving the probe fluctuation characteristics (see SI Sec. 2.F).\par

A GP prior is assigned to the target waveform $E_{\mathrm{THz}}(t)$. The covariance between adjacent temporal samples of the electric field is modeled with a kernel, $G(t, t') = A_c(t)\, A_c(t')\, \exp\!\left( -\frac{(t - t')^2}{2\ell^2} \right)$, where $A_c(t)$ sets the observation window and $\ell$ is the temporal correlation scale between adjacent time pixels $t$ and $t'$. Unlike SVD-based methods~\cite{yellampalle2005algorithm,wu2018electro}, the explicit connection between $E_{THz}$ and prior allows prior parameters to be directly chosen from measurements. Therefore, the correlation scale $\ell$ is determined from the null structure observed in the spectrum of $E_{\mathrm{THz,dist}}$. In the frequency domain, the GP corresponds to a Gaussian spectrum centered at zero frequency and recovers the information loss due to the nulls in the spectrum. To ensure that the prior spans the nulls present in the measured spectrum, the $1~\%$ level of the GP prior wing is placed at the next theoretically predicted null beyond the last observed null. For instance, if the last observed null is $f_{n}=\sqrt{\frac{(2n-1)\alpha}{2\pi}}$, then the next null is $f_{n+1}=\sqrt{\frac{(2n+1)\alpha}{2\pi}}$, and $\ell$ is selected accordingly. This provides a prior-selection rule that does not require a separately measured delay-line reference (SI Sec. 3.D).\par

The GP prior also controls how the inversion behaves near noise-sensitive spectral regions. SVD-based regularization keeps or removes modes according to their singular values, but this criterion cannot distinguish between weak physical modes caused by spectral nulls and modes dominated by noise. Consequently, noise can disturb the regularization and degrade the reconstruction~\cite{hansen1990truncated,hansen1998rank}. In contrast, the explicit GP prior is assigned directly to the THz waveform spectrum. With the explicit prior, weak modes near the nulls are not discarded solely by their scale, but are retained when the prior-supported signal is stronger than the measured noise contribution (SI Sec. 3.D). This makes the framework more robust to systematic noise than SVD-based cutoff methods. As quantified in SI Sec. 3.D and Fig.~S5, the GP reconstruction lowers the time-domain waveform RMSE relative to Tikhonov regularization from 0.160 to 0.0504 at 20 \% noise level. Also, the amplitude ratio between the reconstructed and true spectra is only 0.044 for Tikhonov regularization, while the GP reconstruction maintains a ratio of 1.07 at the first null frequency. Overcoming this reconstruction deficiency in SVD-based methods requires tuning the regularization process, which in turn requires external knowledge of the correct waveform.\par

The posterior is then assembled in this latent space. The likelihood $P_{\mathrm{Like}}(E_{\mathrm{THz,dist}}\mid E_{\mathrm{THz}},\theta)$ compares the observed signal $E_{THz,dist}(\tau)$ with the forward model result $(I_m - I_c)/I_c$, using a Gaussian noise variance measured from shot-to-shot fluctuations (Eq. S28 in SI Sec. 3.B). The prior combines a GP prior on $E_{\mathrm{THz}}(t)$, a super-Gaussian prior on the modulation coefficient $k$ that extends from $0.1-0.2$, and empirical priors on $\mathbf{w}_A$ and $\mathbf{w}_P$. Bayes' theorem gives $P_{\mathrm{Post}}(E_{\mathrm{THz}},\theta\mid E_{\mathrm{THz,dist}})\propto P_{\mathrm{Like}}P_{\mathrm{Prior}}$, the target distribution for inversion.\par

Two algorithms are used to characterize the posterior. MAP estimation finds the posterior mode by minimizing the negative log-posterior with a quasi-Newton BFGS optimizer (SI Eq. S49). A Laplace approximation then represents the posterior locally as a Gaussian centered at the MAP solution, with the covariance estimated from the inverse Hessian of the negative log-posterior. This provides an efficient local estimate of the waveform uncertainty, but it assumes that the posterior is well described by the curvature around a single peak (SI Sec. 3.E). MALA is used as a full-sampling benchmark: it draws samples from the posterior using a drift-diffusion proposal preconditioned by the MAP Hessian, with adaptive step-size control targeting an acceptance rate of $\sim57.4\%$ (SI Sec. 3.F). 

As a calibration case, the two algorithms are compared on an ideal bipolar waveform ($T_{THz} = 0.2~\mathrm{ps}$) with a distorted observable $E_{THz,dist}(\tau)$ that is calculated through a forward model at $\alpha=12.5~\mathrm{ps}^{-2}$ (Fig.~\ref{fig:Fig2}(b),~(c)). Both MAP and MALA recover the true waveform within the 95\% credible interval and show close agreement on the waveform scale, whereas the $E_{THz,dist}(t)$ deviates from the truth with a maximum residual of $\sim1$ around $t = \pm0.5~\mathrm{ps}$. This agreement indicates that the MAP-approximated posterior is sufficient to recover the $E_{THz}(t)$ sets for ideal THz waveforms.\par

\section{Waveform reconstruction results}\label{Waveform Reconstruction Results}
\subsection{Reconstruction comparison between MAP and MALA}
\begin{figure}[!t]
\centering
\includegraphics[width=0.95\linewidth]{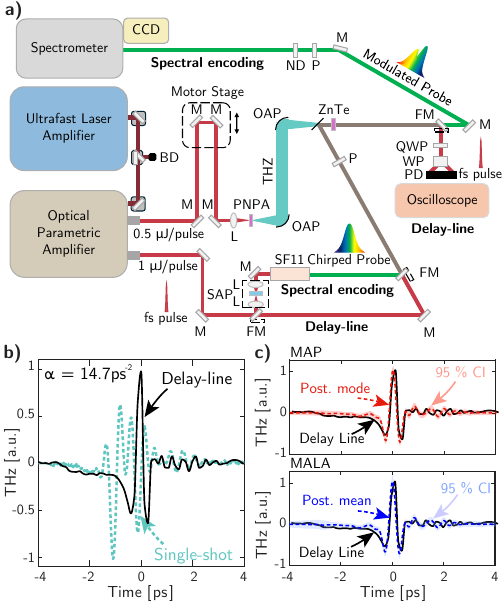}
\caption{(a) Experimental setup of single-shot and delay-line THz-TDS. Beam dump (BD), flip-mirror (FM), lens (L), mirror (M), neutral density filter (ND), off-axis parabolic mirror (OAP), photodetector (PD), polarizer (P), quarter-wave plate (QWP), sapphire (SAP), and Wollaston prism (WP) are included. (b) Raw single-shot (cyan) and delay-line (black) THz waveforms. (c) Inversion result of the single-shot data using the MAP (red) and MALA (blue) algorithms, compared with the delay-line reference (black). The 95\% credible intervals are shown as shaded bands (red for MAP, blue for MALA).}
\label{fig:Fig3}
\end{figure}

To validate the Bayesian workflow experimentally, single-shot THz-TDS measurements were performed under six probe conditions. The single-shot THz-TDS system was built around an amplified femtosecond ytterbium laser (1030~nm, 100~kHz) feeding an optical parametric amplifier, whose synchronized outputs provided the 1532~nm pump for THz generation and the 780~nm probe for electro-optic sampling (Fig.~\ref{fig:Fig3}(a)). THz transients were generated by optical rectification in a 500-$\mu\mathrm{m}$ PNPA crystal. Because spectral encoding distortion depends sensitively on the probe condition, two probe spectra were prepared by omitting (NSP) or inserting (SP) a sapphire stage, and three SF11 glass rods (50, 100, and 150~mm) imposed three chirp rates, yielding six probe configurations ($\alpha = 14.5$--$40~\mathrm{ps}^{-2}$) for evaluating the inversion algorithm. The chirped probe and THz pulse were overlapped on a 2-mm $\langle 110 \rangle$ ZnTe crystal, where the THz-induced Pockels birefringence modulated the probe polarization. This modulation was converted to intensity by crossed polarizers and recorded with a 0.5~m imaging spectrometer and EMCCD camera. A delay-line branch with balanced electro-optic detection was implemented in parallel to provide reference waveforms to validate the reconstructions. By scanning the THz pulse by adjusting the probe delay and measuring the modulation with balanced photodetection, this branch records the conventional time-domain EO waveform. The THz beam path was purged with dry nitrogen to keep the relative humidity below 4\% (see SI Sec. 2).\par
To examine whether the MAP--MALA agreement also holds for experimental data, the two inversion algorithms were applied to the NSP probe with the 150~mm SF11 rod case, which has the lowest $\alpha = 14.52~\mathrm{ps}^{-2}$ and strongest distortion among the six configurations. The raw single-shot waveform deviated from the delay-line reference, particularly around the leading peak from $-2~\mathrm{ps}$ to $0~\mathrm{ps}$, demonstrating oscillatory artifacts (Fig.~\ref{fig:Fig3}(b)). Both MAP and MALA recovered the delay-line waveform within the 95\% credible interval and yielded similar posterior distributions for the waveform variables (Fig.~\ref{fig:Fig3}(c)), suggesting that MAP-based inference is sufficient for experimental waveform reconstruction. Therefore, inversions on the remaining five probe configurations were performed using the MAP estimate (see SI Sec. 4.A).\par

\subsection{Time-domain results across different chirp parameters}\label{Time-domain Results}
A practical reconstruction method must be validated under the probe conditions that are actually required in experiments. In practice, the pulse duration, chirp rate, and spectral profile are adjusted according to the target measurement. For instance, low-frequency THz responses require a long observation window. In this case, strong dispersion with broadened probe bandwidth can be applied to increase both the probe pulse duration time and chirp rate. Therefore, the robustness of the reconstruction framework is validated across different chirp parameters.\par
\begin{figure}[!b]
\centering
\includegraphics[width=0.95\linewidth]{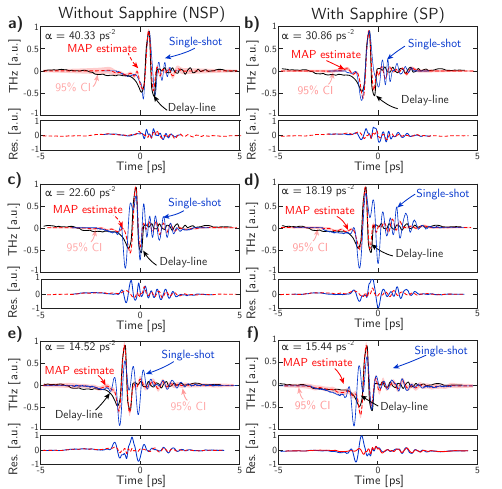}
\caption{Time-domain THz waveform inversion results for the probe without sapphire (NSP, (a), (c), and (e)) and with sapphire (SP, (b), (d), and (f)) for three SF11 rod lengths, 50~mm ((a) and (b)), 100~mm ((c) and (d)), and 150~mm ((e) and (f)). In each panel, the upper subplot shows the MAP estimate (red) overlaid on the delay-line reference (black dashed) and the raw single-shot signal (blue), together with the 95\% credible interval (shaded band). The lower subplot shows the residuals of the MAP estimate (red dashed) and the raw single-shot signal (blue) with respect to the delay-line reference (black). The chirp parameter $\alpha$ is annotated in each panel.}
\label{fig:Fig4}
\end{figure}
Time-domain THz waveform inversion results across the six probe configurations are summarized in Fig.~\ref{fig:Fig4}. For the NSP probe, the chirp parameter decreases from $\alpha=40.33~\mathrm{ps}^{-2}$ at the 50~mm SF11 rod (Fig.~\ref{fig:Fig4}(a)) to $22.60~\mathrm{ps}^{-2}$ at 100~mm (Fig.~\ref{fig:Fig4}) and $14.52~\mathrm{ps}^{-2}$ at 150~mm (Fig.~\ref{fig:Fig4}(e)). For the SP probe, $\alpha$ similarly decreases from $30.86~\mathrm{ps}^{-2}$ at 50~mm (Fig.~\ref{fig:Fig4}(b)) to $18.19~\mathrm{ps}^{-2}$ at 100~mm (Fig.~\ref{fig:Fig4}(d)) and $15.44~\mathrm{ps}^{-2}$ at 150~mm (Fig.~\ref{fig:Fig4}(f)). In each panel, the MAP estimate is shown in red with its 95\% credible interval, the delay-line reference is shown as a black dashed curve, and the raw single-shot waveform is shown in blue. The lower subplot reports the corresponding residuals relative to the delay-line reference.\par

As the SF11 rod length increases, the dispersion temporally stretches the probe field $E_c(t)$ from $2.8~\mathrm{ps}$ to $7.2~\mathrm{ps}$ and reduces the chirp rate $\alpha$. Therefore, the raw single-shot measurements exhibit the distortion described in \eqref{eq:THz-TDS_distortion}, which scales with $1/\alpha^2$. Consequently, the residuals become larger for the longer SF11 rods, particularly in the 150~mm cases for both NSP and SP probes (Fig.~\ref{fig:Fig4}(e),(f)). Furthermore, the residuals are concentrated around the THz peak from $-2~\mathrm{ps}$ to $0~\mathrm{ps}$, where the rapid field variation makes $E'_{\mathrm{THz}}$ and $E''_{\mathrm{THz}}$ large in the distortion term of \eqref{eq:THz-TDS_distortion}. In contrast, the MAP estimate recovers the THz waveform within the 95\% credible interval for all six cases, and its residuals remain below $\sim0.5$, smaller than the raw single-shot residuals, especially over the main pulse window. This consistency across two spectral profiles and three chirp conditions demonstrates that the algorithm recovers the underlying THz waveform despite changes in probe dispersion and spectral shape.\par
\begin{figure}[!b]
\centering
\includegraphics[width=0.95\linewidth]{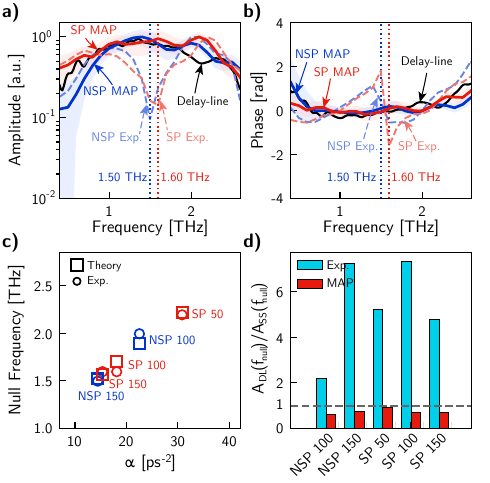}
\caption{Frequency-domain reconstruction results and null frequency analysis across the six cases shown in Fig.~\ref{fig:Fig4}. (a) Amplitude and (b) phase spectra for NSP (blue) and SP (red) with the 150 mm SF11 rod. In each case, the single-shot experimental spectrum (dashed line) exhibits a null frequency that is recovered in the MAP estimate (solid line); the shaded band denotes the 95\% credible interval, and the delay-line reference (black line) is overlaid for comparison. (c) Theoretical (squares) and experimental (circles) null frequencies as a function of $\alpha$ for all six cases. (d) Amplitude ratio $A_{\mathrm{DL}}(f_{\mathrm{null}})/A_{\mathrm{SS}}(f_{\mathrm{null}})$ evaluated at the null frequency $f_{null} = \sqrt{\frac{\alpha}{2\pi}}$ for the single-shot experiment (cyan) and the MAP estimate (red).}
\label{fig:Fig5}
\end{figure}
\subsection{Frequency-domain results across different chirp parameters}\label{Frequency-domain Results}
Additionally, the frequency-domain inversion results and the null frequency analysis are presented in Fig.~\ref{fig:Fig5}. For 150 mm SF11 rod cases (Fig.~\ref{fig:Fig5}(a), (b)), the raw single-shot spectra exhibit amplitude nulls and accompanying phase jumps at 1.50~THz (NSP, $\alpha = 14.51~\mathrm{ps}^{-2}$) and 1.60~THz (SP, $\alpha = 15.44~\mathrm{ps}^{-2}$). On the other hand, the MAP estimates fill in these nulls and recover the amplitude envelope of the delay-line reference within the 95\% credible interval, and the corresponding phase profiles follow the reference without phase jumps.\par

The null-frequency trend across all six cases is summarized in Fig.~\ref{fig:Fig5}(c). The theoretical null frequencies are calculated from $f_{\mathrm{null}}=\sqrt{\alpha/(2\pi)}$ using the mean $\alpha$ of each configuration~\cite{roussel2022phase}. Except for the NSP 50~mm case, whose predicted null lies outside the measured THz spectral range  $\sim2.6~\mathrm{THz}$, the experimental null positions agree with the theoretical values over $\alpha=14.52$--$30.86~\mathrm{ps}^{-2}$, corresponding to null frequencies from 1.5 to 2.2~THz. The degree of recovery is quantified in Fig.~\ref{fig:Fig5}(d) by the amplitude ratio $A_{\mathrm{DL}}(f_{\mathrm{null}})/A_{\mathrm{SS}}(f_{\mathrm{null}})$. The raw single-shot ratios range from 2 to 7, reflecting the depth of the nulls in the distorted spectra. In contrast, the MAP ratios remain near unity across all cases at the same frequencies, indicating that the reconstruction at the null frequency is consistent with the delay-line reference. The corresponding spectra and recovery results for the NSP and SP probes with 50 and 100~mm SF11 rods are provided in SI Sec. 4.C.\par

\section{Discussion}\label{Discussion}
The remaining discrepancy between MAP estimates and delay-line references in Figs.~\ref{fig:Fig4} and~\ref{fig:Fig5} can be interpreted as a combination of systematic mismatch between measurements of the THz waveform and the model. The Bayesian forward model uses a finite empirical parameterization of the chirped probe field, so probe fluctuations outside the retained PCA basis or errors in the calibrated group-delay curve can propagate into the reconstructed waveform.  In addition, the single-shot waveform and the delay-line reference are not direct samples of an identical field, but outputs of different measurement systems, $E_{\mathrm{single\mbox{-}shot}}(t)=\mathcal{H}_{\mathrm{SS}}[E_{\mathrm{THz,dist}}(t)]$ and $E_{\mathrm{delay\mbox{-}line}}(t)=\mathcal{H}_{\mathrm{DL}}[E_{\mathrm{THz}}(t)]$, where $\mathcal{H}_{\mathrm{SS}}$ represents the effective system response of the single-shot branch, including the spectrometer response and alignment-dependent factors. Similarly, $\mathcal{H}_{\mathrm{DL}}$ represents the delay-line branch response, including balanced electro-optic detection and lock-in filtering. The mismatch may also arise from the background birefringence and differences in the effective detection bandwidths of the two branches. Therefore, proper characterization of the probe $E_c(t)$ and spectral encoding system forward model is essential to ensure robust reconstruction reliability in further applications.\par

This work demonstrates that spectral encoding distortion can be corrected across various probe conditions without adding optical hardware. This robustness is essential because the chirped probe needs to be adjusted based on the frequency resolution that is needed in a THz measurement. Such adjustments change the chirp rate and probe spectrum. As a result, null frequencies are shifted, and the encoding response is modified. By validating the reconstruction across six probe conditions spanning two spectral profiles and multiple chirp rates, we show that the Bayesian framework can be applied to a broad class of probe settings.\par

Furthermore, the framework reduces the need for a reference in quantitative measurements because the priors can be constructed from information available within the single-shot measurement and probe characterization, including probe electric field and modulation factor. In addition, the null structure identifies the spectral regions where the encoding response is deficient.  This measured information loss can become a criteria of the THz waveform prior. The same prior-based structure also improves robustness to noise. By introducing an explicit prior form that can be designed to directly access the weak modes near spectral nulls, those modes are recovered through the balance between the prior-supported waveform content and the measured noise level. This allows weak components near the nulls to contribute to the reconstruction while suppressing the noise amplification that can occur in SVD-based methods.\par

More broadly, because the prior and forward model are explicit modeling components, the method is not limited to the GP prior or to THz-TDS. The same formulation can be adapted to other spectral encoding modalities that share the phase loss problem due to the structure of \eqref{eq:general_spectral_encoding_forward}, including chirped-pulse pump-probe spectroscopy, electro-optic spectral decoding, photonic time-stretch measurements, and Raman/CARS-based spectral encoding. For each modality, the probe field, readout kernel, and prior can be parameterized using the corresponding experimental characterization. This enables existing spectrally encoded data to be reinterpreted at the software level and provides a general path toward quantitative single-shot spectroscopy without redesigning the optical system.

\begin{backmatter}
\bmsection{Acknowledgements}
This research is supported by the Office of Naval Research grant N00014-23-1-2306, with Ryan Hoffman as Program Manager.

\bmsection{Disclosures} The authors declare no conflicts of interest.

\bmsection{Data availability} Data underlying the results presented in this paper are not publicly available at this time but may be obtained from the authors upon reasonable request.

\bmsection{Supplementary Information}
See Supplementary Information for supporting content.
\end{backmatter}

% Bibliography
\bibliography{reference}

@article{zhang2025ultra,
  title={Ultra-fast optical time-domain transformation techniques},
  author={Zhang, Yusheng and Tao, Chenning and Luo, Si and Lau, Kuen Yao and Zheng, Jiancheng and Huang, Lin and Zhang, Aiguo and Sheng, Liwen and Ling, Qiang and Guan, Zuguang and others},
  journal={Nature Reviews Methods Primers},
  volume={5},
  number={1},
  pages={11},
  year={2025},
  publisher={Nature Publishing Group UK London}
}

@article{mahjoubfar2017time,
  title={Time stretch and its applications},
  author={Mahjoubfar, Ata and Churkin, Dmitry V and Barland, St{\'e}phane and Broderick, Neil and Turitsyn, Sergei K and Jalali, Bahram},
  journal={Nature Photonics},
  volume={11},
  number={6},
  pages={341--351},
  year={2017},
  publisher={Nature Publishing Group UK London}
}

@article{grace2025single,
  title={Single-shot spatiotemporal plasma density measurements with a chirped probe pulse},
  author={Grace, Elizabeth S and Longman, Andrew and Zeraouli, Ghassan and Maricle, Stephen and Attiyah, Danny and Clark, Jerry and Welch, Ethan and Linder, Austin and Lemos, Nuno and Mariscal, Derek A and others},
  journal={Optica},
  volume={12},
  number={9},
  pages={1522--1528},
  year={2025},
  publisher={Optica Publishing Group}
}

@article{ilyakov2022terahertz,
  title={Terahertz-wave decoding of femtosecond extreme-ultraviolet light pulses},
  author={Ilyakov, Igor and Agarwal, Naman and Deinert, J-C and Liu, Jia and Yaroslavtsev, Alexander and Foglia, Laura and Kurdi, Gabor and Mincigrucci, Riccardo and Principi, Emiliano and Jakob, Gerhard and others},
  journal={Optica},
  volume={9},
  number={5},
  pages={545--550},
  year={2022},
  publisher={Optica Publishing Group}
}

@article{dulat2024single,
  title={Single-shot, spatio-temporal analysis of relativistic plasma optics},
  author={Dulat, Ankit and Lad, Amit D and Aparajit, C and Choudhary, Anandam and Ved, Yash M and Veisz, Laszlo and Ravindra Kumar, G},
  journal={Optica},
  volume={11},
  number={8},
  pages={1077--1084},
  year={2024},
  publisher={Optica Publishing Group}
}

@article{genty2021machine,
  title={Machine learning and applications in ultrafast photonics},
  author={Genty, Go{\"e}ry and Salmela, Lauri and Dudley, John M and Brunner, Daniel and Kokhanovskiy, Alexey and Kobtsev, Sergei and Turitsyn, Sergei K},
  journal={Nature Photonics},
  volume={15},
  number={2},
  pages={91--101},
  year={2021},
  publisher={Nature Publishing Group UK London}
}

@article{yang2016terahertz,
  title={Terahertz multiheterodyne spectroscopy using laser frequency combs},
  author={Yang, Yang and Burghoff, David and Hayton, Darren J and Gao, Jian-Rong and Reno, John L and Hu, Qing},
  journal={Optica},
  volume={3},
  number={5},
  pages={499--502},
  year={2016},
  publisher={Optical Society of America}
}

@article{brown2021bayesian,
  title={Bayesian framework for THz-TDS plasma diagnostics},
  author={Brown, Nathan P and Grauer, Samuel J and Deibel, Jason A and Walker, Mitchell LR and Steinberg, Adam M},
  journal={Optics Express},
  volume={29},
  number={4},
  pages={4887--4901},
  year={2021},
  publisher={Optical Society of America}
}

@article{teo2015invited,
  title={Invited Article: Single-shot THz detection techniques optimized for multidimensional THz spectroscopy},
  author={Teo, Stephanie M and Ofori-Okai, Benjamin K and Werley, Christopher A and Nelson, Keith A},
  journal={Review of scientific instruments},
  volume={86},
  number={5},
  year={2015},
  publisher={AIP Publishing}
}

@article{ofori2025unveiling,
  title={Unveiling structural effects on the DC conductivity of warm dense matter via terahertz spectroscopy and ultrafast electron diffraction},
  author={Ofori-Okai, Benjamin K and Descamps, Adrien and Toro, Edna R and Ikeya, Megan and Hansen, Stephanie B and Mo, Mianzhen and Baczewski, Andrew D and Brown, Danielle and Fletcher, Luke B and McBride, Emma E and others},
  journal={Nature Communications},
  volume={16},
  number={1},
  pages={10541},
  year={2025},
  publisher={Nature Publishing Group UK London}
}

@article{tancin2021ultrafast,
  title={Ultrafast-laser-absorption spectroscopy in the mid-infrared for single-shot, calibration-free temperature and species measurements in low-and high-pressure combustion gases},
  author={Tancin, Ryan J and Goldenstein, Christopher S},
  journal={Optics Express},
  volume={29},
  number={19},
  pages={30140--30154},
  year={2021},
  publisher={Optical Society of America}
}

@book{stanton1994optical,
  title={Optical Generation and Detection of Carriers in Ultrafast Pump-Probe Spectroscopy of Semiconductors},
  author={Stanton, Christopher J and Kuznetsov, Alex V and Kim, Chang Sub},
  booktitle={Coherent Optical Interactions in Semiconductors},
  pages={307--311},
  year={1994},
  publisher={Springer}
}

@article{zewail2000femtochemistry,
  title={Femtochemistry: Atomic-scale dynamics of the chemical bond},
  author={Zewail, Ahmed H},
  journal={The Journal of Physical Chemistry A},
  volume={104},
  number={24},
  pages={5660--5694},
  year={2000},
  publisher={ACS Publications}
}

@article{weakly2023revealing,
  title={Revealing core-valence interactions in solution with femtosecond X-ray pump X-ray probe spectroscopy},
  author={Weakly, Robert B and Liekhus-Schmaltz, Chelsea E and Poulter, Benjamin I and Biasin, Elisa and Alonso-Mori, Roberto and Aquila, Andrew and Boutet, S{\'e}bastien and Fuller, Franklin D and Ho, Phay J and Kroll, Thomas and others},
  journal={Nature Communications},
  volume={14},
  number={1},
  pages={3384},
  year={2023},
  publisher={Nature Publishing Group UK London}
}

@article{adak2019observation,
  title={Observation of ultrafast laser-plasma evolution by pump-probe reflectometry and Doppler spectrometry},
  author={Adak, Amitava and Singh, Prashant Kumar and Lad, Amit D and Chatterjee, Gourab and Kumar, G Ravindra},
  journal={arXiv preprint arXiv:1909.05814},
  year={2019}
}

@article{wang2020single,
  title={Single-shot ultrafast imaging attaining 70 trillion frames per second},
  author={Wang, Peng and Liang, Jinyang and Wang, Lihong V},
  journal={Nature communications},
  volume={11},
  number={1},
  pages={2091},
  year={2020},
  publisher={Nature Publishing Group UK London}
}

@article{wilke2002single,
  title={Single-shot electron-beam bunch length measurements},
  author={Wilke, I. and MacLeod, Allan M. and Gillespie, W. Allan and Berden, G. and Knippels, G. M. H. and van der Meer, A. F. G.},
  journal={Physical Review Letters},
  volume={88},
  number={12},
  pages={124801},
  year={2002},
  publisher={APS}
}

@book{berden2004high,
  title={High temporal resolution, single-shot electron bunch-length measurements},
  author={Berden, Giel and Redlich, B and Van Der Meer, AFG and Jamison, SP and MacLeod, AM and Gillespie, WA},
  booktitle={Free Electron Lasers 2003},
  pages={II--49},
  year={2004},
  publisher={Elsevier}
}

@article{chien2000single,
  title={Single-shot chirped-pulse spectral interferometry used to measure the femtosecond ionization dynamics of air},
  author={Chien, C. Y. and La Fontaine, B. and Desparois, A. and Jiang, Z. and Johnston, T. W. and Kieffer, J. C. and P{\'e}pin, H. and Vidal, F. and Mercure, H. P.},
  journal={Optics Letters},
  volume={25},
  number={8},
  pages={578--580},
  year={2000},
  publisher={Optical Society of America}
}

@article{kim2002single,
  title={Single-shot supercontinuum spectral interferometry},
  author={Kim, K. Y. and Alexeev, I. and Milchberg, H. M.},
  journal={Applied Physics Letters},
  volume={81},
  number={22},
  pages={4124--4126},
  year={2002},
  publisher={American Institute of Physics}
}

@article{patel2020simplified,
  title={Simplified single-shot supercontinuum spectral interferometry},
  author={Patel, Dhruvit and Jang, Dogeun and Hancock, Scott W. and Milchberg, Howard M. and Kim, Ki-Yong},
  journal={Optics Express},
  volume={28},
  number={8},
  pages={11023--11032},
  year={2020},
  publisher={Optical Society of America}
}

@article{polli2010effective,
  title={Effective temporal resolution in pump-probe spectroscopy with strongly chirped pulses},
  author={Polli, Dario and Brida, Daniele and Mukamel, Shaul and Lanzani, Guglielmo and Cerullo, Giulio},
  journal={Physical Review A},
  volume={82},
  number={5},
  pages={053809},
  year={2010},
  publisher={APS}
}

@article{shkrob2004frequency,
  title={Frequency-domain single-shot ultrafast transient absorption spectroscopy using chirped laser pulses},
  author={Shkrob, Ilya A. and Oulianov, Dmitri A. and Crowell, Robert A. and Pommeret, Stanislas},
  journal={Journal of Applied Physics},
  volume={96},
  number={1},
  pages={25--33},
  year={2004},
  publisher={American Institute of Physics}
}

@article{knutsen2006chirped,
  title={Chirped coherent anti-Stokes Raman scattering for high spectral resolution spectroscopy and chemically selective imaging},
  author={Knutsen, Kelly P and Messer, Benjamin M and Onorato, Robert M and Saykally, Richard J},
  journal={The Journal of Physical Chemistry B},
  volume={110},
  number={12},
  pages={5854--5864},
  year={2006},
  publisher={ACS Publications}
}

@article{couture2023single,
  title={Single-pulse terahertz spectroscopy monitoring sub-millisecond time dynamics at a rate of 50 kHz},
  author={Couture, Nicolas and Cui, Wei and Lippl, Markus and Ostic, Rachel and Fandio, D{\'e}fi Junior Jubgang and Yalavarthi, Eeswar Kumar and Vishnuradhan, Aswin and Gamouras, Angela and Joly, Nicolas Y and M{\'e}nard, Jean-Michel},
  journal={Nature Communications},
  volume={14},
  number={1},
  pages={2595},
  year={2023},
  publisher={Nature Publishing Group UK London}
}

@article{yellampalle2005algorithm,
  title={Algorithm for high-resolution single-shot THz measurement using in-line spectral interferometry with chirped pulses},
  author={Yellampalle, B and Kim, KY and Rodriguez, George and Glownia, JH and Taylor, Antoinette Jane},
  journal={Applied Physics Letters},
  volume={87},
  number={21},
  year={2005},
  publisher={AIP Publishing}
}

@article{wu2018electro,
  title={Electro-optic sampling of optical pulses and electron bunches for a compact THz-FEL source},
  author={Wu, Bang and Zhang, Zhe and Cao, Lei and Fu, Qiang and Xiong, Yongqian},
  journal={Infrared Physics \& Technology},
  volume={92},
  pages={287--294},
  year={2018},
  publisher={Elsevier}
}

@book{hansen1998rank,
  title={Rank-deficient and discrete ill-posed problems: numerical aspects of linear inversion},
  author={Hansen, Per Christian},
  year={1998},
  publisher={SIAM}
}

@article{hansen1990truncated,
  title={Truncated singular value decomposition solutions to discrete ill-posed problems with ill-determined numerical rank},
  author={Hansen, Per Christian},
  journal={SIAM Journal on Scientific and Statistical Computing},
  volume={11},
  number={3},
  pages={503--518},
  year={1990},
  publisher={SIAM}
}

@article{sun1998analysis,
	author = {Sun, FG and Jiang, Zhiping and Zhang, X-C},
	journal = {Applied Physics Letters},
	number = {16},
	pages = {2233--2235},
	publisher = {American Institute of Physics},
	title = {Analysis of terahertz pulse measurement with a chirped probe beam},
	volume = {73},
	year = {1998}
}

@article{jiang2000measurement,
	author = {Jiang, Zhiping and Zhang, Xi-Cheng},
	journal = {IEEE journal of quantum electronics},
	number = {10},
	pages = {1214--1222},
	publisher = {IEEE},
	title = {Measurement of spatio-temporal terahertz field distribution by using chirped pulse technology},
	volume = {36},
	year = {2000}
}

@article{roussel2022phase,
	author = {Roussel, El{\'e}onore and Szwaj, Christophe and Evain, Cl{\'e}ment and Steffen, Bernd and Gerth, Christopher and Jalali, Bahram and Bielawski, Serge},
	journal = {Light: Science \& Applications},
	number = {1},
	pages = {14},
	publisher = {Nature Publishing Group UK London},
	title = {Phase Diversity Electro-optic Sampling: A new approach to single-shot terahertz waveform recording},
	volume = {11},
	year = {2022}
}

@article{nallapareddy2025characterization,
	author = {Nallapareddy, Charan R and Underwood, Thomas C},
	doi = {10.1063/5.0288545},
	journal = {APL Photonics},
	number = {11},
	pages = {116105},
	publisher = {AIP Publishing},
	title = {Characterization and control of signal distortion in chirped pulse single-shot terahertz detection},
	volume = {10},
	year = {2025}
}

@article{Nallapareddy2025a,
	author = {Nallapareddy, C.R. and Underwood, T.C.},
	doi = {10.1038/s41467-025-60550-6},
	journal = {Nature Communications},
	pages = {5188},
	title = {{Quantitative single-shot Supercontinuum-Enhanced Terahertz Spectroscopy (SETS)}},
	volume = {16},
	year = {2025}
}

@article{nallapareddy2026probing,
  title={Probing hysteresis and bifurcation dynamics in reactive radio frequency plasmas},
  author={Nallapareddy, Charan R and Saha, Avijit and Hood-McFadden, Drue and Underwood, Thomas C},
  journal={Applied Physics Letters},
  volume={128},
  number={9},
  year={2026},
  publisher={AIP Publishing}
}

@book{saha2026traceable,
  title={Traceable Bayesian Uncertainty Quantification in Single-Shot Terahertz Spectroscopy of Plasmas},
  author={Saha, Avijit and Kang, Minsoo and R. Nallapareddy, Charan and C. Underwood, Thomas},
  booktitle={AIAA SCITECH 2026 Forum},
  pages={1229},
  year={2026},
  Publisher={American Institute of Aeronautics and Astronautics}
}

@article{kang2026robust,
  title={Robust Bayesian parameter estimation for spectral encoded single-shot THz spectroscopy},
  author={Kang, Minsoo and Saha, Avijit and Nallapareddy, Charan R and Underwood, Thomas C},
  journal={Optics Express},
  volume={34},
  number={12},
  pages={21587--21616},
  year={2026},
  publisher={Optica Publishing Group}
}

@article{matlis2010single,
  title={Single-shot spatiotemporal measurements of ultrashort THz waveforms using temporal electric-field cross correlation},
  author={Matlis, NH and Plateau, GR and van Tilborg, Jeroen and Leemans, WP},
  journal={Journal of the Optical Society of America B},
  volume={28},
  number={1},
  pages={23--27},
  year={2010},
  publisher={Optical Society of America}
}

@book{kaipio2005statistical,
  title={Statistical and Computational Inverse Problems},
  author={Kaipio, Jari and Somersalo, Erkki},
  publisher={Springer},
  address={Dordrecht},
  year={2005},
  doi={10.1007/b138659}
}

@article{stuart2010inverse,
  title={Inverse problems: A Bayesian perspective},
  author={Stuart, Andrew M.},
  journal={Acta Numerica},
  volume={19},
  pages={451--559},
  year={2010},
  doi={10.1017/S0962492910000061}
}
\end{document}